\documentclass[epj]{svjour}
\usepackage{graphics}
\usepackage[utf8]{inputenc}
\usepackage{graphicx}
\usepackage{psfrag}
\usepackage{amssymb}
\usepackage{amsmath}
\usepackage{epstopdf}
\usepackage{color}
\usepackage{bm}
\begin{document}
\title{Strong decays of the $\Xi(1620)$ as a $\Lambda\bar{K}$ and $\Sigma\bar{K}$ molecule}
\author{Yin Huang\inst{1}
\and Lisheng Geng\inst{2,3,4,5}\thanks{\emph{Corresponding author:} lisheng.geng@buaa.edu.cn}
%
}                     
%
%
\institute{School of Physical Science and Technology, Southwest Jiaotong University, Chengdu 610031,China
\and School of Physics, Beihang University, Beijing 102206, China
\and Beijing Key Laboratory of Advanced Nuclear Materials and Physics, Beihang University, Beijing 102206, China
\and Beijing Advanced Innovation Center for Big Data-based Precision Medicine, Beihang University, Beijing100191, China
\and School of Physics and Microelectronics, Zhengzhou University, Zhengzhou, Henan 450001, China}
\date{Received: date / Revised version: date}
%
\abstract{
In this work, we study the strong decays of the newly observed $\Xi(1620)^0$ assuming that
it is a meson-baryon molecular state of $\Lambda\bar{K}$ and $\Sigma\bar{K}$.
We consider four possible spin-parity assignments $J^P=1/2^{\pm}$ and $3/2^{\pm}$ for
the $\Xi(1620)^0$, and evaluate its partial decay width into $\Xi\pi$ via hadronic loops
with the help of effective Lagrangians.  In comparison with the Belle data, the calculated
decay width favors the spin-party assignment $1/2^-$ while the other spin-parity assignments
do not yield a decay width consistent with data in the molecule picture.   We find that about
$52\%$-$68\%$ of the total width comes from the $\bar{K}\Lambda$ channel, while the rest is provided by
the $\bar{K}\Sigma$ channel. As a result, both channels are important in explaining the strong decay of
the $\Xi(1620)^0$. This information is helpful to further understand the nature of the $\Xi(1620)^0$.
} 
\maketitle

\section{Introduction}\label{sec1}
Understanding  baryon spectroscopy  and searching for missing  baryon resonances  are hot
topics in hadron physics.  From the viewpoint of the quark model, the number of $\Xi$ states should
be comparable with that of nucleon resonances.  At present, there are eleven $\Xi$ baryons listed
in the review of the Particle Data Group (PDG)~\cite{Tanabashi:2018oca}, which is far less than
the number of nucleon baryons.    Among them, the $\Xi(1620)$, $\Xi(2120)$, and $\Xi(2500)$
are three peculiar states, since their are catalogued in the PDG with only one star and they
spin and parity are unknown~\cite{Tanabashi:2018oca}.  In other words, the experimental evidence
for these three $\Xi$ bayrons are quit poor, and it is not yet clear whether
they really exist.  Fortunately, the  $\Xi(1620)$ was recently observed in the $\Xi^{-}\pi^{+}$ final state by the Belle Collaboration~\cite{Sumihama:2018moz}.  Its mass and width are, respectively,
\begin{align}
&M=1610.4\pm{6.0}(stat)^{+5.9}_{-3.5}(syst)~~ {\rm MeV},\nonumber\\
&\Gamma=59.9\pm4.8(stat)^{+2.8}_{-3.0}(syst)~~{\rm MeV},
\end{align}
which are consistent with the earlier measured values~\cite{Ross:1972bf,Briefel:1977bp}.
Its spin-parity, however,  remains unknown.

It needs to be stressed that the quark model, originally pioneered by  Gell-Mann and Zweig,
still remains a useful yardstick for baryon spectroscopy.  However, one common characteristic of the quark model is
that it is very difficult to accommodate the $\Xi(1620)$~\cite{Capstick:1986bm,Blask:1990ez}.  In particular,
the low mass of the $\Xi(1620)$ is puzzling in the quark model if its existence is further confirmed by future experiments.
It is very interesting to note that the authors of Ref.~\cite{Azimov:2003bb} try to assign the $\Xi(1620)$ as a
conventional $uss$ or $dss$ state with $J^P=1/2^{-}$.   Although their model satisfies the Gell-Mann-Okubo
mass relation, it requires the existence of  very low mass nucleon and $\Lambda$ resonances, which have not been discovered yet.

These peculiar properties of the $\Xi(1620)$ can be naturally accounted for in the hadronic molecule picture.  Indeed, in
Ref.~\cite{Ramos:2002xh}, Ramos et al.suggested to identify the $\Xi(1620)$ as a dynamically generated $S$-wave $\Xi$
resonance based on an unitary extension of chiral perturbation theory, which predicts a $\Xi$ resonance with a mass
around 1606 MeV.   In other similar approaches (that differ in details)~\cite{GarciaRecio:2003ks,Gamermann:2011mq,Miyahara:2016yyh},
the $\Xi(1620)$ is also dynamically generated, with a relatively larger decay width. This state strongly couples to $\pi\Xi$ and $\bar{K}\Lambda$,
and it is thought to originate from the strong attraction in the $\pi\Xi$ channel~~\cite{GarciaRecio:2003ks,Gamermann:2011mq,Miyahara:2016yyh}.
In the Skyrme model~\cite{Oh:2007cr}.  Yongseok predicted two $\Xi$ resonances with $J^P=1/2^{-}$, which have masses consistent
with those of the $\Xi(1620)$ and $\Xi(1690)$.

Following the discovery of the $\Xi$(1620),  several theoretical studies have been performed~\cite{Wang:2019krq,Chen:2019uvv}.
In the Bethe-Salpeter equation approach under the ladder and instantaneous approximations,  based on the analysis of the
mass spectrum and the two-body strong decays, the $\Xi$(1620) is explained as $\bar{K}\Lambda$ or $\Sigma{}\bar{K}$
bound states with spin-parity $J^P = 1/2^-$~\cite{Wang:2019krq}.  We note that the decay widths are 36.94 MeV
and 9.35 MeV for the $\bar{K}\Lambda$ and $\Sigma{}\bar{K}$ bound states, respectively.   In comparison with the Belle data,
it is obvious that the $\Xi$(1620) has a larger contribution from the $\bar{K}\Lambda$ component than the $\Sigma{}\bar{K}$ component.
From this perspective, it is easy to understand the results of
 Ref.~\cite{Chen:2019uvv}, where it was shown that the $\bar{K}\Lambda$ interaction is strong enough to form a $\Xi$ bound state with a mass about 1620 MeV and $J^P=1/2^{-}$ in the framework
of the One-Boson-Exchange (OBE) model.
\begin{figure*}[htbp]
\begin{center}
\includegraphics[scale=0.65]{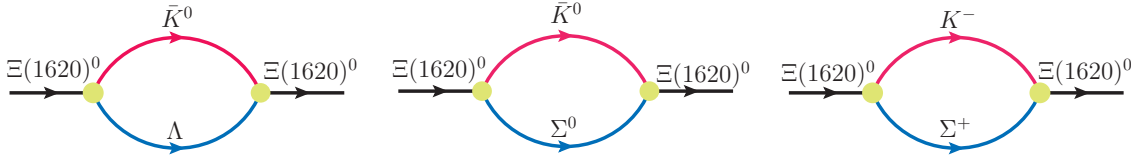}
\end{center}
\caption{Self-energy of the $\Xi$(1620) state.}\label{msef}
\end{figure*}

Although the studies of Refs.~\cite{Ramos:2002xh,GarciaRecio:2003ks,Gamermann:2011mq,Miyahara:2016yyh,Wang:2019krq,Chen:2019uvv}
seem to indicate that the $\Xi(1620)^0$ is a hadronic molecule state,  more theoretical efforts are
needed to fully understand its nature.  Considering both
the theoretical results~\cite{GarciaRecio:2003ks,Gamermann:2011mq,Miyahara:2016yyh,Chen:2019uvv} and
the latest experimental measurement that the mass of $\Xi(1620)^{0}$  is about 3 MeV below the $\bar{K}^{0}\Lambda$ threshold ~\cite{Tanabashi:2018oca}, it is reasonable to regard $\Xi(1620)^{0}$ as a bound
state of $\bar{K}^{0}\Lambda$.   Note  that  in Refs.~\cite{Ramos:2002xh,Wang:2019krq} the $\Xi(1620)$ is treated as a
meson-baryon state with large $\Lambda\bar{K}$ and $\Sigma\bar{K}$ components.  In the present work we study
 the $\Xi\pi$ decay mode of the $\Xi$(1620),  using an effective Lagrangian approach and assuming that the $\Xi$(1620)
is a hadronic molecular state of $\Lambda\bar{K}$ and $\Sigma\bar{K}$ with the following four spin-parity assignments: $J^P=1/2^{\pm}$ and $3/2^{\pm}$.

This work is organized as follows. The theoretical formalism is explained in Sec.~\ref{section:2}.
The predicted partial decay width is presented in Sec.~\ref{section:3}, followed by a short summary
in the last section.

\section{FORMALISM AND INGREDIENTS}\label{section:2}
In order to calculate the strong decay width, $\Xi(1620)\to\Xi\pi$, in the molecular scenario with
different spin-parity assignments for the $\Xi(1620)$, we first need to compute the couplings with its
components $\bar{K}\Lambda$ and $\bar{K}\Sigma$ via the loop diagrams shown in Fig.~(\ref{msef}).

The simplest effective Lagrangian describing the $\Xi^{*}\bar{K}Y$ coupling can be expressed
as~\cite{Huang:2018bed,Dong:2010gu}
\begin{align}
{\cal{L}}^{1/2^{\pm}}_{\Xi^{*}}(x)&=g_{\Xi^{*}\bar{K}Y}\int{}d^4y\Phi(y^2)\bar{K}(x+\omega_{Y}y)\Gamma\nonumber\\
                                  &\times{}Y(x-\omega_{\bar{K}}y)\bar{\Xi}^{*}(x)\label{eq1},\\
{\cal{L}}^{3/2^{\pm}}_{\Xi^{*}}(x)&=g_{\Xi^{*}\bar{K}Y}\int{}d^4y\Phi(y^2)\bar{K}(x+\omega_{Y}y)\Gamma\nonumber\\
                        &\times\partial_{\mu}Y(x-\omega_{K}y)\bar{\Xi}^{*\mu}(x)\label{eq2},
\end{align}
where $Y$ denotes either $\Lambda$ or $\Sigma$ (for an isovector baryon $Y$, $Y$ should be replaced with $\vec{Y}\cdot\vec{\tau}$, where
$\tau$ is the isospin matrix),
$\omega_{\bar{K}}=m_{\bar{K}}/(m_{\bar{K}}+m_{Y})$, $\omega_{Y}=m_{Y}/(m_{\bar{K}}+m_{Y})$, and $\Gamma$ is
the corresponding Dirac matrix reflecting the spin-parity of the $\Xi(1620)$.   Here $\Gamma=\gamma^{5}$ for $J^{p}=1/2^{+}$ and $3/2^{-}$, while for $J^{p}=1/2^{-}$ and $3/2^{+}$, $\Gamma=1$.  In the above Lagrangian, an effective
correlation function $\Phi(y^2)$ is  introduced not only to describe the distribution of the constituents,  $\bar{K}$
and $Y$, in the hadronic molecular $\Xi(1620)$ state but also to make the Feynmann diagrams ultraviolet finite, which is often chosen to be of the  following form~\cite{Huang:2018bed,Dong:2010gu,Dong:2014ksa,Dong:2010xv,Dong:2009tg,Faessler:2007gv,Faessler:2007us,Dong:2008gb,Dong:2009uf,Dong:2009yp,Dong:2017rmg,Dong:2014zka,Dong:2013kta,Dong:2013iqa,Dong:2013rsa,Dong:2012hc,Dong:2011ys,Dong:2017gaw} ,
\begin{align}
\Phi(p_E^2)\doteq\exp(-p_E^2/\beta^2)
\end{align}
with $p_E$ being the Euclidean Jacobi momentum and $\beta$ being the size parameter which characterizes the distribution
of the components inside the molecule.  At present, the value of $\beta$ still could not be accurately determined from
first principles, therefore it should better be determined by experimental data.  The experimental total widths of some states that
can be considered as molecules~\cite{Huang:2018bed,Dong:2010gu,Dong:2014ksa,Dong:2010xv,Dong:2009tg,Faessler:2007gv,Faessler:2007us,Dong:2008gb,Dong:2009uf,Dong:2009yp,Dong:2017rmg,Dong:2014zka,Dong:2013kta,Dong:2013iqa,Dong:2013rsa,Dong:2012hc,Dong:2011ys,Dong:2017gaw} can be well explained with $\beta=1.0$ GeV.   Therefore we take $\beta=1.0$
GeV in this work to study whether the $\Xi(1620)$ can be interpreted as a molecule composed of $\bar{K}\Lambda$ and $\bar{K}\Sigma$.

\begin{figure*}[htbp]
\begin{center}
\includegraphics[scale=0.55]{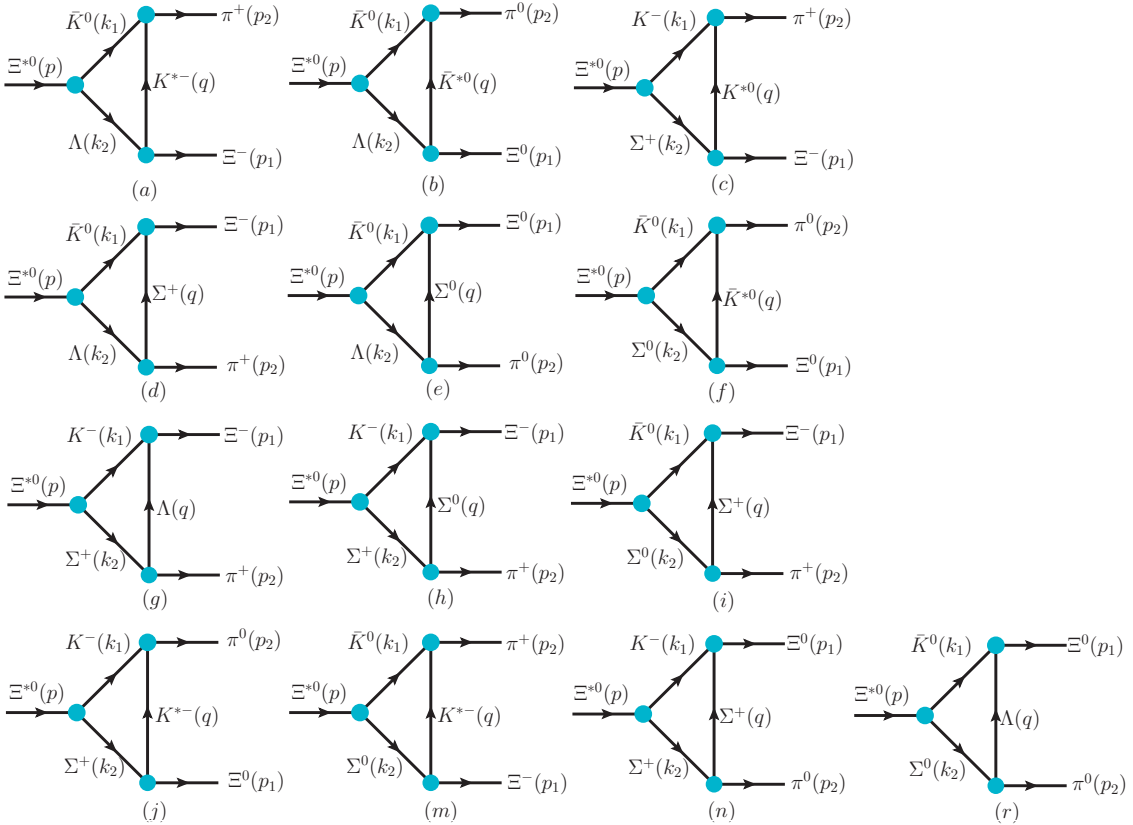}
\caption{Feynman diagrams for the $\Xi^{*0}\to{}\pi\Xi$ decay processes.  We also show
the definitions of the kinematics ($p,k_1,k_2,p_1,p_2$, and $q$) used in the calculation.}\label{fety}
\end{center}
\end{figure*}
With the effective Lagrangians in Eq.~(\ref{eq1}) and Eq.~(\ref{eq2}), we can compute the Feynmann diagrams shown in Fig.~(\ref{msef}),
and obtain the self-energy of the $\Xi(1620)$,
\begin{align}
\Sigma^{1/2}_{\Xi^{*}}(k_0)&=\sum_{Y=\Lambda,\Sigma^0,\Sigma^{+}}{\cal{C}}_{Y}g^2_{\Xi^{*}\bar{K}Y}\int\frac{d^4k_1}{(2\pi)^4}\Phi^2[(k_1-k_0\omega_{Y})_E^2]\nonumber\\
                             &\times\Gamma\frac{k\!\!\!/_1+m_{Y}}{k_1^2-m^2_{Y}}\Gamma\frac{1}{(k_1-k_0)^2-m^2_{\bar{K}}}\label{eqn1},\\
\Sigma^{\mu\nu3/2}_{\Xi^{*}}(k_0)&=\sum_{Y=\Lambda,\Sigma^0,\Sigma^{+}}{\cal{C}}_{Y}g^2_{\Xi^{*}\bar{K}Y}\int\frac{d^4k_1}{(2\pi)^4}\Phi^2[(k_1-k_0\omega_{Y})_E^2]\nonumber\\
                             &\times\Gamma\frac{k\!\!\!/_1+m_{Y}}{k_1^2-m^2_{Y}}\Gamma\frac{1}{(k_1-k_0)^2-m^2_{\bar{K}}}\}k_1^{\mu}k_1^{\nu},\label{eqn2}
\end{align}
where $k_0^2=m^2_{\Xi^{*}}$ with $k_0, m_{\Xi^{*}}$ denoting the four momenta and the mass of the $\Xi(1620)$, respectively,
$k_1$, $m_{\bar{K}}$, and $m_{Y}$ are the four-momenta, the mass of the $\bar{K}$ meson, and the mass of the $Y$ baryon, respectively.
Here, we set $m_{\Xi^{*}}=m_{Y}+m_{\bar{K}}-E_b$ with $E_b$ being the binding energy of $\Xi(1620)^{0}$.  Isospin symmetry implies that
\begin{align}
{\cal{C}}_Y=\left\{
\begin{aligned}
1           &  & {Y=\Lambda} \\
1/3         &  & {Y=\Sigma^0} \\
2/3         &  & {Y=\Sigma^+}.
\end{aligned}
\right.
\end{align}
The coupling constant $g_{\Xi^{*}\bar{K}Y}$ is determined by the compositeness condition~\cite{Dong:2010gu,Dong:2014ksa,Dong:2010xv,Dong:2009tg,Faessler:2007gv,Faessler:2007us,Dong:2008gb,Dong:2009uf,Dong:2009yp,Dong:2017rmg,Dong:2014zka,Dong:2013kta,Dong:2013iqa,Dong:2013rsa,Dong:2012hc,Dong:2011ys,Dong:2017gaw}. It implies that the renormalization constant of the hadron wave function is set to zero, i.e.,
\begin{align}
Z_{\Xi^{*}}=x_{\bar{K}\Sigma}+x_{\bar{K}\Lambda}-\frac{d\Sigma^{1/2(3/2-T)}_{\Xi^{*}}}{dk\!\!\!/_0}|_{k\!\!\!/_0=m_{\Xi^{*}}}=0\label{eqn3},
\end{align}
where  $x_{AB}$ is the probability to find the $\Xi(1620)^0$ in the hadronic state $AB$ with normalization $x_{\bar{K}\Sigma}+x_{\bar{K}\Lambda}=1.0$. The $\Sigma^{3/2-T}_{\Xi_b^{*}}$ is the transverse part of the self-energy operator $\Sigma^{\mu\nu3/2}_{\Xi^{*}}$, related to $\Sigma^{\mu\nu3/2}_{\Xi^{*}}$ via
\begin{align}
\Sigma^{\mu\nu3/2}_{\Xi^{*}}(k_0)=(g_{\mu\nu}-\frac{k_0^{\mu}k_0^{\nu}}{k_0^2})\Sigma^{3/2-T}_{\Xi^{*}}+\cdots.\label{eqn4}
\end{align}

For the $\Xi(1620)$, because of phase space, only the  strong decay into $\Xi\pi$  and radiative decay are allowed.
However, radiative decay widths are often in the keV regime and are far less than their strong counterparts.  Therefore,
in the present work, we focus on the $\pi\Xi$ two body decay of the $\Xi(1620)$ in the $\bar{K}\Lambda-\bar{K}\Sigma$ molecular picture mediated by the exchange of
$\bar{K}^{*}$, $\Lambda$, and $\Sigma$.  The corresponding Feynmann diagrams are shown in Fig.~(\ref{fety}).

To evaluate the diagrams, in addition to the Lagrangians in Eq.~(\ref{eq1}) and Eq.~(\ref{eq2}), the following effective Lagrangian, responsible for the interactions of light pseudoscalar and vector mesons are needed as well~\cite{Garzon:2012np}
\begin{align}
{\cal{L}}_{VPP}&=-ig\langle[P,\partial_{\mu}P]V^{\mu}\rangle\label{eqw9},
\end{align}
where $P$ and $V^{\mu}$ are the $SU(3)$ pseudoscalar and vector meson matrices, respectively, and $\langle...\rangle$ denotes
 trace in the flavor space.  The meson matrices are~\cite{Garzon:2012np}
\begin{equation}
P=
\left(
  \begin{array}{ccc}
    \frac{\pi^{0}}{\sqrt{2}}+\frac{\eta}{\sqrt{6}} &    \pi^{+}                                        &       K^{+}\\
    \pi^{-}                                       &    -\frac{\pi^{0}}{\sqrt{2}}+\frac{\eta}{\sqrt{6}} &       K^{0}\\
    K^{-}                                         &    \bar{K}^{0}                                     &       -\frac{2}{\sqrt{6}}\eta
  \end{array}
\right)\label{eq7}.
\end{equation}
and
\begin{equation}
V_{\mu}=
\left(
  \begin{array}{cccc}
    \frac{1}{\sqrt{2}}(\rho^{0}+\omega) & \rho^{+}                             &  K^{*+}      \\
    \rho^{-}                            & \frac{1}{\sqrt{2}}(-\rho^{0}+\omega) &  K^{*0}       \\
     K^{*-}                             & \bar{K}^{*0}                         &  \phi       \\
  \end{array}
\right)_{\mu}.
\end{equation}
The coupling $g$ is fixed from the strong decay width of $K^{*}\to{}K\pi$.  With the help of Eq.~(\ref{eqw9}), the two-body decay width
 $\Gamma(K^{*+}\to{}K^{0}\pi^{+})$ is related to $g$ as
 \begin{align}
 \Gamma(K^{*+}\to{}K^{0}\pi^{+})=\frac{g^2}{6\pi{}m^2_{K^{*+}}}{\cal{P}}^3_{\pi{}K^{*}}=\frac{2}{3}\Gamma_{K^{*+}},
 \end{align}
where ${\cal{P}}_{\pi{}K^{*}}$ is the three-momentum of the $\pi$ in the rest frame of the  $K^{*}$.
Using the experimental strong decay width, $\Gamma_{K^{*+}}=50.3\pm0.8$ MeV, and the masses of the particles listed in Table.~\ref{table1}, we obtain $g=4.64$~\cite{Tanabashi:2018oca}.
\begin{table}[h!]
\centering
\caption{Masses of the particles needed in the present work (in units of MeV).}\label{table1}
\begin{tabular}{ccccccccc}
\hline\hline
   $\Lambda$    &  $\Xi^{-}$    &  $\Xi^{0}$   & $K^{0}$     &$\pi^{0}$                        \\
   $1115.683$   &  $1321.71$    &  $1314.86$   & $497.611$   &$134.977$                        \\ \hline
   $K^{*0}$     &  $K^{*\pm}$   &  $K^{\pm}$   & $\pi^{\pm}$ &$\Sigma^{+}$  &$\Sigma^{0}$      \\
   $895.55$     &  $891.76$     &  $493.68$    & $139.57$    &$1189.37$     &$1192.642$        \\  \hline  \hline
\end{tabular}
\end{table}

Moreover, meson-baryon interactions are also needed and can be obtained from the following chiral Lagrangians~\cite{Garzon:2012np}
\begin{align}
{\cal{L}}_{VBB}&=g(\langle\bar{B}\gamma_{\mu}[V^{\mu},B]\rangle+\langle\bar{B}\gamma_{\mu}B\rangle\langle{}V^{\mu}\rangle)\label{vbb},\\
{\cal{L}}_{PBB}&=\frac{F}{2}\langle\bar{B}\gamma_{\mu}\gamma_5[u^{\mu},B]\rangle+\frac{D}{2}\langle\bar{B}\gamma_{\mu}\gamma_5\{u^{\mu},B\}\rangle\label{Pbb},
\end{align}
where $F=0.51$, $D=0.75$~\cite{Garzon:2012np,Borasoy:1998pe} and at the lowest order $u^{\mu}=-\sqrt{2}\partial^{\mu}P/f$ with $f=93$ MeV, and $B$ is the $SU(3)$ matrix of the baryon octet
\begin{equation}
B=
\left(
  \begin{array}{ccc}
    \frac{1}{\sqrt{2}}\Sigma^{0}+\frac{1}{\sqrt{6}}\Lambda  & \Sigma^{+}                                               &  p      \\
     \Sigma^{-}                                             & -\frac{1}{\sqrt{2}}\Sigma^{0}+\frac{1}{\sqrt{6}}\Lambda  &  n       \\
     \Xi^{-}                                                & \Xi^{0}                                                  &  -\frac{2}{\sqrt{6}}\Lambda     \\
  \end{array}
\right).
\end{equation}

Putting all the pieces together, we obtain the following strong decay amplitudes,
\begin{align}
{\cal{M}}_{a}&(\Xi^{*0}\to\Xi^{-}\pi^{+})=-(i)^3\frac{3}{\sqrt{6}}g^2g_{\Xi^{*}\Lambda\bar{K}}\int\frac{d^4q}{(2\pi)^4}\nonumber\\
                         &\times{}\Phi[(k_1\omega_{\Lambda}-k_2\omega_{\bar{K}^{0}})^2]\bar{u}(p_1)\gamma_{\mu}\frac{k\!\!\!/_2+m_{\Lambda}}{k_2^2-m^2_{\Lambda}}\nonumber\\
                         &\times\Gamma{}\{u(p),ik_2^{\rho}u_{\rho}(p)\}\frac{1}{k_1^2-m^2_{\bar{K}^{0}}}(k_1^{\nu}+p^{\nu}_{2})\nonumber\\
                         &\times\frac{-g^{\mu\nu}+q^{\mu}q^{\nu}/m^2_{K^{*-}}}{q^{2}-m^2_{K^{*-}}},\\
{\cal{M}}_{b}&(\Xi^{*0}\to\Xi^{0}\pi^{0})=(i)^3\frac{\sqrt{3}}{2}g^2g_{\Xi^{*}\Lambda\bar{K}}\int\frac{d^4q}{(2\pi)^4}\nonumber\\
                         &\times{}\Phi[(k_1\omega_{\Lambda}-k_2\omega_{\bar{K}^{0}})^2]\bar{u}(p_1)\gamma_{\mu}\frac{k\!\!\!/_2+m_{\Lambda}}{k_2^2-m^2_{\Lambda}}\nonumber\\
                         &\times\Gamma{}\{u(p),ik_2^{\rho}u_{\rho}(p)\}\frac{1}{k_1^2-m^2_{\bar{K}^{0}}}(k_1^{\nu}+p^{\nu}_{2})\nonumber\\
                         &\times\frac{-g^{\mu\nu}+q^{\mu}q^{\nu}/m^2_{K^{*0}}}{q^{2}-m^2_{K^{*0}}},\\
{\cal{M}}_{c}&(\Xi^{*0}\to\Xi^{-}\pi^{+})=(i)^3\sqrt{\frac{2}{3}}g^2g_{K^{-}\Sigma^{+}\Xi^{*}}\int\frac{d^4q}{(2\pi)^4}\nonumber\\
                         &\times\Phi[(k_1\omega_{\Sigma^{+}}-k_2\omega_{K^{-}})^2]\bar{u}(p_1)\gamma_{\mu}\frac{k\!\!\!/_2+m_{\Sigma^{+}}}{k_2^2-m^2_{\Sigma^{+}}}\nonumber\\
                         &\times\Gamma{}\{u(p),ik_2^{\rho}u_{\rho}(p)\}\frac{1}{k_1^2-m^2_{K^{-}}}(k_1^{\nu}+p^{\nu}_{2})\nonumber
\end{align}
\begin{align}
                         &\times\frac{-g^{\mu\nu}+q^{\mu}q^{\nu}/m^2_{K^{*0}}}{q^{2}-m^2_{K^{*0}}},\\
{\cal{M}}_{d}&(\Xi^{*0}\to\Xi^{-}\pi^{+})=(i)^3\frac{D(D+F)}{\sqrt{6}f^2}g_{\Xi^{*}\Lambda\bar{K}}\int\frac{d^4q}{(2\pi)^4}\nonumber\\
                   &\times{}\Phi[(k_1\omega_{\Lambda}-k_2\omega_{\bar{K}^{0}})^2]\bar{u}(p_1)k\!\!\!/_1\gamma_5\frac{q\!\!\!/+m_{\Sigma^{+}}}{q^2-m^2_{\Sigma^{+}}}p\!\!\!/_2\nonumber\\
                   &\times{}\gamma_5\frac{k\!\!\!/_2+m_{\Lambda}}{k_2^2-m^2_{\Lambda}}\Gamma{}\{u(p),ik_2^{\rho}u_{\rho}(p)\}\frac{1}{k_1^2-m^2_{\bar{K}^0}},\\
{\cal{M}}_{e}&(\Xi^{*0}\to\Xi^{0}\pi^{0})=-(i)^3\frac{D(D+F)}{\sqrt{6}f^2}g_{\Xi^{*}\Lambda\bar{K}}\int\frac{d^4q}{(2\pi)^4}\nonumber\\
                   &\times{}\Phi[(k_1\omega_{\Lambda}-k_2\omega_{\bar{K}^{0}})^2]\bar{u}(p_1)k\!\!\!/_1\gamma_5\frac{q\!\!\!/+m_{\Sigma^{0}}}{q^2-m^2_{\Sigma^{0}}}p\!\!\!/_2\nonumber\\
                   &\times{}\gamma_5\frac{k\!\!\!/_2+m_{\Lambda}}{k_2^2-m^2_{\Lambda}}\Gamma{}\{u(p),ik_2^{\rho}u_{\rho}(p)\}\frac{1}{k_1^2-m^2_{\bar{K}^0}},\\
{\cal{M}}_{f}&(\Xi^{*0}\to\Xi^{0}\pi^{0})=(i)^3\frac{1}{2\sqrt{3}}g^2g_{\Xi^{*}\Sigma^0\bar{K}^0}\int\frac{d^4q}{(2\pi)^4}\nonumber\\
                   &\times{}\Phi[(k_1\omega_{\Sigma^0}-k_2\omega_{\bar{K}^{0}})^2]\bar{u}(p_1)\gamma_{\mu}\frac{k\!\!\!/_2+m_{\Sigma^{0}}}{k_2^2-m^2_{\Sigma^{0}}}\nonumber\\
                   &\times\Gamma{}\{u(p),ik_2^{\rho}u_{\rho}(p)\}\frac{1}{k_1^2-m^2_{\bar{K}^{0}}}(k_1^{\nu}+p^{\nu}_{2})\nonumber\\
                   &\times\frac{-g^{\mu\nu}+q^{\mu}q^{\nu}/m^2_{K^{*0}}}{q^{2}-m^2_{K^{*0}}},\\
{\cal{M}}_{g}&(\Xi^{*0}\to\Xi^{-}\pi^{+})=-(i)^3\frac{D(D-3F)}{3\sqrt{6}f^2}g_{\Xi^{*}\Sigma^{+}K^-}\int\frac{d^4q}{(2\pi)^4}\nonumber\\
                   &\times{}\Phi[(k_1\omega_{\Sigma^{+}}-k_2\omega_{{K}^{-}})^2]\bar{u}(p_1)k\!\!\!/_1\gamma_5\frac{q\!\!\!/+m_{\Lambda}}{q^2-m^2_{\Lambda}}p\!\!\!/_2\nonumber\\
                   &\times{}\gamma_5\frac{k\!\!\!/_2+m_{\Sigma^{+}}}{k_2^2-m^2_{\Sigma^{+}}}\Gamma{}\{u(p),ik_2^{\rho}u_{\rho}(p)\}\frac{1}{k_1^2-m^2_{K^-}},\\
{\cal{M}}_{h}&(\Xi^{*0}\to\Xi^{-}\pi^{+})=-(i)^3\frac{F(D+F)}{2f^2}g_{\Xi^{*}\Sigma^{+}K^-}\int\frac{d^4q}{(2\pi)^4}\nonumber\\
                   &\times{}\Phi[(k_1\omega_{\Sigma^{+}}-k_2\omega_{{K}^{-}})^2]\bar{u}(p_1)k\!\!\!/_1\gamma_5\frac{q\!\!\!/+m_{\Sigma^{0}}}{q^2-m^2_{\Sigma^{0}}}p\!\!\!/_2\nonumber\\
                   &\times{}\gamma_5\frac{k\!\!\!/_2+m_{\Sigma^{+}}}{k_2^2-m^2_{\Sigma^{+}}}\Gamma{}\{u(p),ik_2^{\rho}u_{\rho}(p)\}\frac{1}{k_1^2-m^2_{K^-}},\\
{\cal{M}}_{i}&(\Xi^{*0}\to\Xi^{-}\pi^{+})=-(i)^3\frac{(D+F)F}{\sqrt{6}f^2}g_{\Xi^{*}\Sigma^0\bar{K}^{0}}\int\frac{d^4q}{(2\pi)^4}\nonumber\\
             &\times{}\Phi[(k_1\omega_{\Sigma^0}-k_2\omega_{\bar{K}^{0}})^2]\bar{u}(p_1)k\!\!\!/_1\gamma_5\frac{q\!\!\!/+m_{\Sigma^{-}}}{q^2-m^2_{\Sigma^{-}}}p\!\!\!/_2\nonumber\\
             &\times{}\gamma_5\frac{k\!\!\!/_2+m_{\Sigma^{0}}}{k_2^2-m^2_{\Sigma^{0}}}\Gamma{}\{u(p),ik_2^{\rho}u_{\rho}(p)\}\frac{1}{k_1^2-m^2_{\bar{K}^0}},\\
{\cal{M}}_{j}&(\Xi^{*0}\to\Xi^{0}\pi^{0})=(i)^3\frac{1}{\sqrt{3}}g^2g_{\Xi^{*}\Sigma^+{K}^{-}}\int\frac{d^4q}{(2\pi)^4}\nonumber\\
             &\times{}\Phi[(k_1\omega_{\Sigma^+}-k_2\omega_{K^{-}})^2]\bar{u}(p_1)\gamma_{\mu}\frac{k\!\!\!/_2+m_{\Sigma^+}}{k_2^2-m^2_{\Sigma^+}}\nonumber\\
             &\times\Gamma{}\{u(p),ik_2^{\rho}u_{\rho}(p)\}\frac{1}{k_1^2-m^2_{{K}^{-}}}(k_1^{\nu}+p^{\nu}_{2})\nonumber\\
             &\times\frac{-g^{\mu\nu}+q^{\mu}q^{\nu}/m^2_{K^{*-}}}{q^{2}-m^2_{K^{*-}}},
\end{align}
\begin{align}
{\cal{M}}_{m}&(\Xi^{*0}\to\Xi^{-}\pi^{+})=(i)^3\frac{1}{\sqrt{6}}g^2g_{\Xi^{*}\Sigma^0\bar{K}^0}\int\frac{d^4q}{(2\pi)^4}\nonumber\\
                   &\times{}\Phi[(k_1\omega_{\Sigma^0}-k_2\omega_{\bar{K}^{0}})^2]\bar{u}(p_1)\gamma_{\mu}\frac{k\!\!\!/_2+m_{\Sigma^{0}}}{k_2^2-m^2_{\Sigma^{0}}}\nonumber\\
                   &\times\Gamma{}\{u(p),ik_2^{\rho}u_{\rho}(p)\}\frac{1}{k_1^2-m^2_{\bar{K}^{0}}}(k_1^{\nu}+p^{\nu}_{2})\nonumber\\
                   &\times\frac{-g^{\mu\nu}+q^{\mu}q^{\nu}/m^2_{K^{*-}}}{q^{2}-m^2_{K^{*-}}},\\
{\cal{M}}_{n}&(\Xi^{*0}\to\Xi^{0}\pi^{0})=-(i)^3\frac{F(D+F)}{\sqrt{3}f^2}g_{\Xi^{*}\Sigma^{+}K^-}\int\frac{d^4q}{(2\pi)^4}\nonumber\\
                   &\times{}\Phi[(k_1\omega_{\Sigma^{+}}-k_2\omega_{{K}^{-}})^2]\bar{u}(p_1)k\!\!\!/_1\gamma_5\frac{q\!\!\!/+m_{\Sigma^{+}}}{q^2-m^2_{\Sigma^{+}}}p\!\!\!/_2\nonumber\\
                   &\times{}\gamma_5\frac{k\!\!\!/_2+m_{\Sigma^{+}}}{k_2^2-m^2_{\Sigma^{+}}}\Gamma{}\{u(p),ik_2^{\rho}u_{\rho}(p)\}\frac{1}{k_1^2-m^2_{K^-}},\\
{\cal{M}}_{r}&(\Xi^{*0}\to\Xi^{-}\pi^{+})=(i)^3\frac{(D-3F)D}{6\sqrt{3}f^2}g_{\Xi^{*}\Sigma^0\bar{K}^{0}}\int\frac{d^4q}{(2\pi)^4}\nonumber\\
             &\times{}\Phi[(k_1\omega_{\Sigma^0}-k_2\omega_{\bar{K}^{0}})^2]\bar{u}(p_1)k\!\!\!/_1\gamma_5\frac{q\!\!\!/+m_{\Lambda}}{q^2-m^2_{\Lambda}}p\!\!\!/_2\nonumber\\
             &\times{}\gamma_5\frac{k\!\!\!/_2+m_{\Sigma^{0}}}{k_2^2-m^2_{\Sigma^{0}}}\Gamma{}\{u(p),ik_2^{\rho}u_{\rho}(p)\}\frac{1}{k_1^2-m^2_{\bar{K}^0}}.
\end{align}
where $\{u(p)$, and $ik_2^{\rho}u_{\rho}(p)\}$ are $J^P=1/2$ and $J^P=3/2$ $\Xi(1620)$ fields,respectively.

Once the amplitudes are determined, the corresponding partial decay width can be easily obtained, which reads as,
\begin{align}
\Gamma(\Xi(1620)^{0}\to\pi^{0}\Xi^{0},\pi^{+}\Xi^{-})=\frac{1}{2J+1}\frac{1}{8\pi}\frac{|\vec{p}_1|}{m^2_{\Xi^{*0}}}\bar{|{\cal{M}}|^2},
\end{align}
where  $J$ is the total angular momentum of the $\Xi(1620)$,  $|\vec{p}_1|$ is the three-momenta of the decay products in the center
of mass frame, the overline indicates the sum over the polarization vectors of the final hadrons.
The total decay width of the $\Xi(1620)^0$ is the sum of $\Gamma(\Xi(1620)^{0}\to\pi^{0}\Xi^{0})$ and $\Gamma(\Xi(1620)^{0}\to\pi^{+}\Xi^{-})$.

\section{RESULTS AND DISCUSSIONS}\label{section:3}
\begin{figure}[htbp]
\begin{center}
\includegraphics[bb=30 00 780 410, clip, scale=0.38]{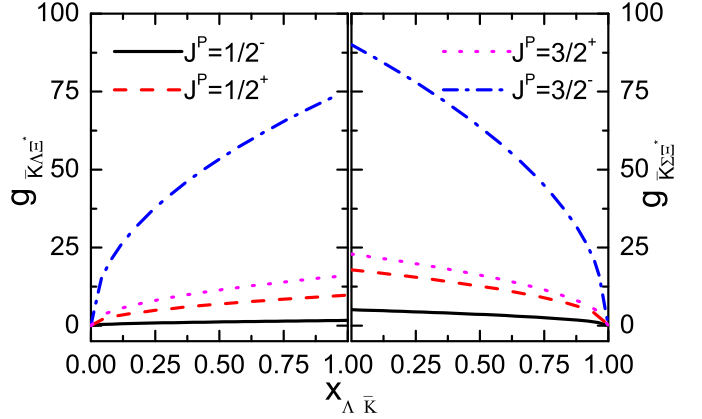}
\caption{Coupling constants of the $\Xi(1620)$ with different $J^P$ assignments as a function of the parameter
$x_{\Lambda\bar{K}}$ which is the probability to find the $\Xi(1620)^0$ in the hadronic component $\bar{K}\Lambda$.}\label{kfety}
\end{center}
\end{figure}

Before calculating the two body decay width, we need to determine the coupling constants relevant to
the effective Lagrangians listed in Eq.~(\ref{eq1}) and Eq.~(\ref{eq2}).  Considering $\Xi(1620)^{0}$ as a $\bar{K}\Lambda-\bar{K}\Sigma$
hadronic molecule, the coupling constants $g_{\bar{K}\Xi^{*}\Lambda}$ and $g_{\bar{K}\Xi^{*}\Sigma}$ can be estimated from
the compositeness condition that we introduced in the previous section.  The $x_{\bar{K}\Lambda}$ dependence of the coupling constants $g_{\bar{K}\Xi^{*}\Lambda}$ and $g_{\bar{K}\Xi^{*}\Sigma}$ are presented in Fig.~\ref{kfety}.
The coupling $g_{\bar{K}\Xi^{*}\Lambda}$ monotonously increases with increasing
$x_{\Lambda\bar{K}}$, and the dependence on $x_{\Lambda\bar{K}}$ is the weakest  for the $J^P =1/2^{-}$ case,
where the $\Xi^{*}(1620)$ is an $S$-wave $\bar{K}\Lambda-\bar{K}\Sigma$ molecular state, while it is
  the strongest for the $J^p=3/2^-$ case.
Comparing $g_{\bar{K}\Xi^{*}\Sigma}$ with $g_{\bar{K}\Xi^{*}\Lambda}$, we find that
their line shapes are very different, i.e.,  the coupling constant $g_{\bar{K}\Xi^{*}\Sigma}$
decreases with increasing $x_{\Lambda\bar{K}}$. We find that $g_{\bar{K}\Xi^{*}\Sigma}$
is the largest for the $J^P=3/2^{-}$ case, is intermediate for the $J^P=1/2^+$ and $J^P=3/2^+$ cases,  and is the smallest for the $J^P=1/2^{-}$ case.
The opposite trend can be easily understood, as the coupling constants $g_{\bar{K}\Xi^{*}\Lambda}$ and $g_{\bar{K}\Xi^{*}\Sigma}$
are directly proportional to the corresponding molecular compositions~\cite{Dong:2009uf}.
Moreover, the relations between  the coupling constants $g_{\bar{K}\Xi^{*}\Sigma}$ and $g_{\bar{K}\Xi^{*}\Lambda}$
can be deduced from Eq.~(\ref{eqn3}) and are given in Table ~\ref{table2}.
\begin{table}[htbp!]
\centering
\caption{Relations between the coupling constants $g_{\bar{K}\Xi^{*}\Sigma}$ and $g_{\bar{K}\Xi^{*}\Lambda}$.}\label{table2}
\begin{tabular}{ccccccccccccccccccccc}
\hline\hline
                        &\multicolumn{2}{c}{$(Ag_{\bar{K}\Xi^{*}\Lambda})^2=1-(Bg_{\bar{K}\Xi^{*}\Sigma})^2$}~~~~~ \\\cline{2-3}
~~~~~spin-parity        &~~~~~ $A$             &$B$             \\\hline
~~~~~$J^P=1/2^{-}$      &~~~~~0.6014           &0.1969             \\
~~~~~$J^P=1/2^{+}$      &~~~~~0.1023           &0.0559             \\
~~~~~$J^P=3/2^{-}$      &~~~~~0.0132           &0.0111             \\
~~~~~$J^P=3/2^{+}$      &~~~~~0.0620           &0.0436             \\
\hline\hline
\end{tabular}
\end{table}

\begin{figure}[htbp]
\begin{center}
\includegraphics[bb=15 20 750 500, clip, scale=0.36]{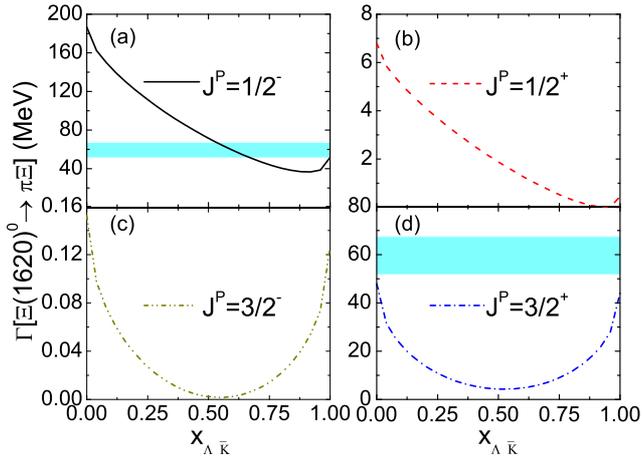}
\caption{Partial decay width of the $\Xi(1620)^0\to\pi\Xi$ with different $J^P$ assignments
depending on the parameter $X_{\bar{K}\Lambda}$.  The black solid, red dashed, dark yellow dash-dot-dotted,
and blue dash-dotted lines stand for the $J^P=1/2^{-}$, $J^P=1/2^{+}$, $J^P=3/2^{-}$, and $J^P=3/2^{+}$ cases,
respectively.  The LT Cyan  bands correspond to the total experimental decay width. }\label{kfety12}
\end{center}
\end{figure}

With the obtained couplings $g_{\bar{K}\Xi^{*}\Lambda}$ and $g_{\bar{K}\Xi^{*}\Sigma}$, the partial decay width of the $\Xi(1620)^0$
can be calculated straightforwardly.   The dependence of the partial decay width on
$X_{\bar{K}\Lambda}$  of the $\Xi(1620)$ for various quantum numbers is given in Fig.~\ref{kfety12}.  In the present study,
we vary $X_{\bar{K}\Lambda}$ from 0.0 to 1.0.   For small $X_{\bar{K}\Lambda}$, the total decay width decreases with
increasing $X_{\bar{K}\Lambda}$.
However, it increases when  $X_{\bar{K}\Lambda}$ varies from 0.91 to 1.00 and 0.54 to 1.00 for the
$J^P = 1/2^{\pm}$ and $J^P = 3/2^{\pm}$ assignments, respectively.

As shown in Fig.~\ref{kfety12}, the LT Cyan bands in these plots denote the experimental data.
For the $J^P=1/2^{+}$ case,  the predicted total decay width increases from 0.011 MeV to 6.787 MeV and is much smaller
than the experimental total width,  which disfavors such a spin-parity assignment for the $\Xi(1620)^0$ in the $\bar{K}\Lambda-\bar{K}\Sigma$
molecular picture.   For the $J^{P}=3/2^{-}$ assignment,  the total decay width is also much smaller than the experimental total width.
This disfavors the assignment of this state as a $\bar{K}\Lambda-\bar{K}\Sigma$ molecular state as well.  The same is also true for the
 $J^P=3/2^{+}$ case.
Hence, only the assignment as an $S-$wave $\bar{K}\Lambda-\bar{K}\Sigma$ molecular state  for the $\Xi(1620)^{0}$
is consistent with the Belle data~\cite{Sumihama:2018moz} when $X_{\bar{K}\Lambda}$ is in the range of 0.52-0.68.
In this region, the total decay width for this state is predicted to be about $50.39-68.79$ MeV.    From  Fig.~\ref{kfety12}
we conclude that the total experimental decay width can be well reproduced, which provides direct evidence that the
observed $\Xi(1620)^{0}$ is an $S-$ wave $\bar{K}\Lambda-\bar{K}\Sigma$ molecular state.

From Fig.~\ref{kfety12}, we also note that the decay width of the observed $\Xi(1620)^{0}$ can not be well reproduced in
a pure $\bar{K}\Lambda$ or pure $\bar{K}\Sigma$ molecular state picture.  Namely, the interference among the two channels is sizable,
leading to a total decay width consistent with the experimental data in the case of the $J^P=1/2^{-}$.  In other
words, the $\bar{K}\Lambda$ channel strongly couples to the $\bar{K}\Sigma$ channel.   However, a completely different
conclusion was drawn in Ref.~\cite{Ramos:2002xh} that there is no interaction between $\bar{K}\Lambda$ and
$\bar{K}\Sigma$.   Comparing our results with those in Ref.~\cite{Chen:2019uvv}, it seems a study of the spectroscopy alone
 does not give a complete picture of its nature.  Furthermore, the $\bar{K}\Lambda$ component provides the dominant
contribution to the partial decay width of the $\pi\Xi$ two-body channel.  This is consistent with the result of
Ref.~\cite{Wang:2019krq} that the $\Xi$(1620) has a larger contribution from the $\bar{K}\Lambda$ channel
than the $\Sigma{}\bar{K}$ channel.

It should be noted that Ref.~\cite{Ramos:2002xh}
showed that the $\Xi(1620)$ couples strongly to the $\pi\Xi$ and the $\bar{K}\Lambda$ channels but very weakly to $\eta\Xi$ and
$\Sigma{}\bar{K}$.  If we treat the $\Xi(1620)$ as a pure $\eta\Xi$ molecular state, we find that the $\eta\Xi$ channel provides
a negligible contribution to the partial decay width into $\pi\Xi$.  Furthermore, based on the Weinberg-Salam compositeness condition,
we find the $\pi\Xi$ component  is significantly suppressed and it contribute negligibly to the partial decay width into $\pi\Xi$.
A possible explanation for this may be that the threshold for $\pi\Xi$ is too low to allow for a bound state at 1620
MeV~\cite{Dong:2013iqa}.   Because of these reasons, the $\eta\Xi$ and $\pi\Xi$ channels are not considered in this work.

\section{Summary}\label{sec:summary}
We have studied the strong decay of the newly observed $\Xi(1620)^{0}$  into $\pi\Xi$ with different spin-parity assignments and assuming that it is a $\bar{K}\Lambda-\bar{K}\Sigma$ molecular state.  With the coupling constants between the $\Xi(1620)^{0}$ and its components determined by the compositeness condition,  we calculated its partial decay width into $\pi\Xi$  via triangle diagrams in an effective Lagrangian approach.  In such a picture, the decay into $\pi\Xi$ occurs by exchanging $K^{*}$, $\Lambda$, and $\Sigma$.   We found that the total decay width can be reproduced with the assumption that the $\Xi(1620)^{0}$ is an $S-$wave $\bar{K}\Lambda-\bar{K}\Sigma$ bound state with $J^P=1/2^{-}$, while the $P-$ and $D-$wave assignments are excluded.  The $\bar{K}\Lambda$ component provides the dominant contribution to the partial decay width into $\pi\Xi$. More specifically,  about $52\%$-$68\%$ of the total decay width is from the $\bar{K}\Lambda$ channel, while the $\bar{K}\Sigma$ channel provides the rest.

\section*{Acknowledgments}
This work is partly supported by the development and Exchange Platform for Theoretic Physics of
Southwest Jiaotong University in 2020(Grants No.11947404).  This work is also supported by the
Fundamental Research Funds for the Central Universities(Grants No. A0920502052001-240).
 We acknowledge the supported by the National Natural Science Foundation of China under
 Grants No.11975041 and No.11735003.

\end{document}